\documentclass[preprint,apl,notitlepage]{revtex4-1}

\usepackage{graphicx}
\usepackage{amssymb}
\usepackage{amsfonts}
\usepackage{amsmath}
\usepackage{amsbsy}
\usepackage{natbib}

\usepackage{comment}
\usepackage{color}
\usepackage[dvipsnames]{xcolor}
\usepackage{subcaption}
\usepackage{hyperref}
\usepackage{cleveref}
\usepackage{changes}
\usepackage{epstopdf}
\renewcommand{\vec}[1]{\mbox{\boldmath$\mathrm{#1}$}}
\def\ind#1{{_{\mathrm{#1}}}}
\newcommand{\be}{\begin{equation}}
\newcommand{\ee}{\end{equation}}
\newcommand{\ben}{\begin{eqnarray}}
\newcommand{\een}{\end{eqnarray}}
\newcommand{\dd}{\mathrm{d}}

\begin{document} 
\title{Ultrafast nanoscale magnetic switching via intense picosecond electron bunches}

\begin{abstract}
	The  magnetic field associated with a picosecond intense electron pulse is shown to  switch  locally the magnetization of extended films and nanostructures and to ignite locally  spin waves excitations. Also topologically protected magnetic textures such as skyrmions can be imprinted swiftly in a sample with a residual Dzyaloshinskii-Moriya  spin-orbital coupling. 
	Characteristics  of the created excitations such as the topological charge or the width of the magnon spectrum can be steered via the duration and the strength of the electron pulses.
	The study points to a possible way for a spatiotemporally controlled generation of magnetic and skyrmionic excitations.
\end{abstract}
\author{A.~F.~Sch\"{a}ffer$^{1,*}$, H.~A.~D\"urr$^2$, J.~Berakdar$^1$}
\affiliation{$^1$Institut f\"ur Physik, Martin-Luther-Universit\"at Halle-Wittenberg, 06099 Halle (Saale), Germany}
\affiliation{$^2$Stanford Institute for Materials and Energy Sciences, SLAC National Accelerator Laboratory, 2575 Sand Hill Road, Menlo Park, California 94025, USA}

\date{\today}
\maketitle

\section{INTRODUCTION}
\label{sect:intro}  
Ultrafast electron microscopy, the introduction of temporal resolution to conventional 3-dimensional electron microscopy, has  opened up the space-time exploration of as diverse as nanomaterials and biostructures \cite{38}. Novel electron optical concepts promise a tremendous increase in temporal and spatial resolution allowing researchers to address new scientific challenges \cite{1,2,3,4,32}. The use of relativistic MeV electron energies is an attractive avenue to deal with the issue of Coulomb repulsion in ultrashort electron pulses bringing single-shot electron microscopy on nm length- and ps time-scales within reach \cite{39,39b}. Electrons with MeV energy travel close to the speed of light. This reduces dramatically the velocity mismatch between the electromagnetic pump and electron probe pulses in time-resolved diffraction experiments \cite{ued_ref}. This has proven indispensable in probing ultrafast processes in fields ranging from gas-phase photochemistry \cite{40} over lattice transformation in low-dimensional systems \cite{41} to the excitation of transient phonons \cite{42}. 

While the use of electron pulses as structural dynamics probes is well established, their use as initiators of dynamical processes has remained almost completely unexplored. In an early study, Tudosa \textit{et. al} used the electromagnetic fields surrounding intense relativistic electron pulses to permanently switch the magnetization direction of ferromagnetic films used in magnetic data storage applications \cite{27}. Here we show that nanofocused relativistic electron pulses provide a unique tool to drive and control magnetization dynamics in nanostructures relevant for spintronics applications \cite{27,43}. The paper is organized as follows. After this introduction, we will describe how our model calculations are able to reproduce the experimental results of Tudosa \textit{et. al} \cite{27}. We will then show how nanofocused electron beams are able to induce switching and magnon excitations in nanowires and can alter the chirality of skyrmions in nanodots. The paper finishes with conclusions and a brief outlook to possible experimental realizations.

Very recently we published a related work, which deals not only with nanostructured materials but also with extended thin films, possessing Dzyaloshinskii-Moriya interaction (DMI)\cite{APL}. There, the creation of skyrmions aside from geometrical confinement is discussed based on the same methods like in this manuscript and is recommended for further reading.   
\section{Results and Discussion}

\subsection{Electron beam induced magnetic switching}
\subsubsection{Methods}
For the purpose of investigating the dynamics of thin magnetic films, micromagnetic simulations solving the Landau-Lifshitz-Gilbert equation (LLG)\cite{34,35} are executed. 
The different magnetic systems are discretized into cuboid lattices, so that each simulation cell is associated with a magnetization $\vec{m}_i=\vec{M}_i/M_{\mathrm{S}}$ normalized to the saturation magnetization $M_{\mathrm{S}}$.
The magnetization dynamics is governed by the  LLG 
\begin{equation}
\dot{\vec{m}}_i =  - \frac{\gamma}{1+\alpha^2}\left\{\vec{m}_i\times \vec{H}_i^{\mathrm{eff}}(t) + \alpha \left[\vec{m}_i\times \left(\vec{m}_i\times \vec{H}_i^{\mathrm{eff}}(t)\right)\right]\right\}\ .
\end{equation}
Here $\gamma=1.76 \cdot 10^{11}$~1/(Ts) denotes the gyromagnetic ratio and $\alpha$ is the dimensionless Gilbert damping parameter.
The  local effective magnetic field 
$\vec{H}_i^{\mathrm{eff}}(t)$ can be calculated following the equation $\mu_0\vec{H}_i^{\mathrm{eff}}(t)=-1/M\ind{S}\delta F/(\delta \vec{m}_i)$ 
and is therefore a functional of the system's total free energy $F=F\ind{EXCH} + F\ind{MCA} + F\ind{DMF} + F\ind{ZMN}+F\ind{DMI}$.
This quantity is influenced  by the exchange interaction of adjacent magnetic moments  $F\ind{EXCH}= - A/c^2\sum_{<ij>}\vec{m}_i \cdot \vec{m}_j$, the magnetocrystalline anisotropy $F\ind{MCA}$, the demagnetizing fields $F\ind{DMF}$, the Zeeman-energy $F\ind{ZMN}$ and the DMI term $F\ind{DMI}$. Further details on the single contributions can be found for example in ref.\cite{36,37}. 
In order to simulate the magnetization dynamics an adaptive Heun solver method has been used. As the excitations of the systems take place in a very short time scale the time step is fixed to $1\,$fs during the time evolution. Using the simulation package \texttt{mumax3}\cite{37} full GPU-based micromagnetic calculations are employed to account for the effect of demagnetizing fields efficiently. 

\begin{figure}
	\begin{center}
		\includegraphics[width=0.5\linewidth]{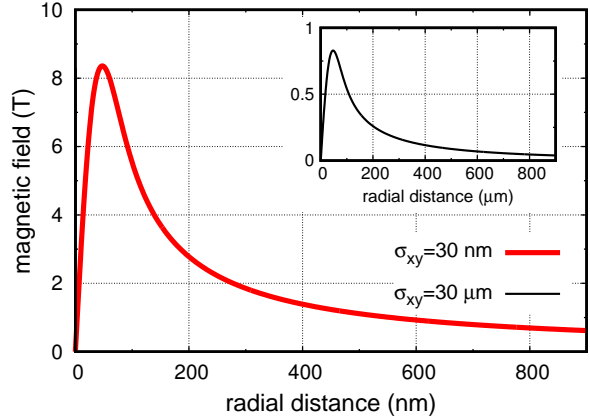}
		\caption{Radial dependence of the peak magnetic field for a pulse duration of $\sigma_t=2.3\,$ps. The red curve corresponds to $\sigma_{xy}=30\,$nm and $N_e=10^8$, whereas the black line relates to $\sigma_{xy}=30\,\mu$m and $N_e=10^{10}$.}
		\label{fig_bfield}
	\end{center}
\end{figure}

As the external magnetic fields, generated by short electron pulses are the main driving mechanism via the Zeeman-coupling, the calculation of these Oersted-like fields is required. The near-field of the electron bunch becomes important for the nanostructures treated in sections \ref{sec:2.2} and \ref{sec:2.3}.
We assume a pulse of electrons with a Gaussian envelope in space and in time or accordingly the propagation direction $j_z(r,\varphi,z)= \frac{N_e e v}{(2\pi)^{3/2}\sigma_{xy}^2\sigma_{z}}\exp\left[-\frac{1}{2}(r/\sigma_{xy})^2-\frac{1}{2}((z-t v)/\sigma_z)^2\right]\ .$
The parameter $N_e$ corresponds to the number of electrons, $e$ is the electron charge, $v$ the average velocity and $\sigma_{xy}$ and $\sigma_z$ are the standard deviations in the related directions. The field's profile resulting from Biot-Savart's law $\vec{B}(\vec{r})=\frac{\mu_0}{4\pi}\int_V \vec{j}(\vec{r}')\times\frac{\vec{r}-\vec{r}'}{|\vec{r}-\vec{r}'|^3}\dd V'$\ ,
is shown in fig. \ref{fig_bfield} for two different sets of parameters. Because of the electron packet propagating in $z$-direction, the magnetic field consists of the $B_\varphi$-component only as inferred from Biot-Savart's law in cylindrical coordinates. The curves shapes are almost the same, as in both cases the radial extension of the beam is much smaller than the standard deviation in the propagation direction. If this is changed the profiles do change as well. The peak field strength for the $30\,\mu$m beam is ten times smaller than the other pulse. This can be deduced from the number of electrons being a hundred times larger, but contrary the beam width is a thousand times wider, which leads to a resulting factor of 10. 
For subsequent calculations, the numerical results for the magnetic field are fitted with a model function, which is dependent on both the beam's standard deviation and the included number of electrons, which are experimentally well accessible parameters.
\subsubsection{Verification of the model}
\begin{figure}[!th]
	\begin{center}
		\includegraphics[width=.5\linewidth]{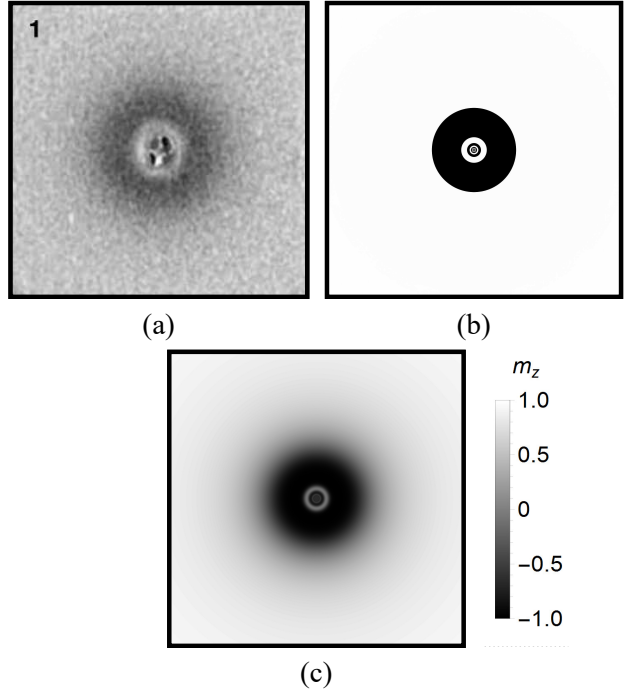}
		\caption{Comparison between experimental (a)\cite{27},   and numerical results (b), (c).
			Both numerical simulations and the experimental data cover an area of $150\times 150\,\mu$m$^2$. In contrast to the panel (b),  in (c) the demagnetizing fields are included in simulations.
			The gray shading  signals the magnetization's $z$-component with white color  meaning $m_z=+\hat{e}_z$ and black  $m_z=-\hat{e}_z$. The $N_e = 10^{10}$ electrons in the beam impinging normal to the sample have an energy of 28\,GeV. The pulse's time-envelope is taken as a Gaussian  with a pulse duration of $\sigma_t = 2.3\,$ps, whereas $\sigma_{xy}=30\,\mu$m. The generated Oersted field has an equivalent time-dependence. }
		\label{fig_comp}
	\end{center}
\end{figure}

In the past years, multiple experiments exploring the field of magnetic switching triggered by fast electrons have been established. One pioneering example can be found in ref.\cite{27}. In this experiment, short electron bunches are shot through thin films (thickness $\approx$14\,nm) of granular CoCrPt-type, in order to explore the ultimate speed limit for precessional magnetic dynamics. The electron pulses have a duration of $\sigma_t=2.3\,$ps a spatial extent of $\sigma_{xy}=30\,\mu$m and included a number of $N_e=10^{10}$ electrons each. After the irradiation, the magnetic domain structure is analyzed, which reveals several domain-wall rings with out-of-plane orientation (\cref{fig_comp}(a)). Starting from a homogeneously magnetized state in $\pm z$ direction the magnetic moments precess around the $\hat{e}_\rho$ unit vector during the pulse and relax either up or down afterward. 
Most of the materials used for magnetic data storage media possesses uniaxial magnetocrystalline anisotropy. CoCrPt-alloys are no exclusion, with an easy axis in the out-of-plane direction, which coincides with the cylindrical $z$-axis. The material specific parameters are chosen to be equal to the measured ones \cite{27}, meaning a  saturation magnetization $M\ind{sat}=517.25\,$kA/m, the uniaxial anisotropy $K\ind{u}=156.98\,\text{kJ}/\text{m}^3$ and the Gilbert damping parameter $\alpha=0.3$. 
Similar to the experimental work, we also consider a thin film of CoCrPt and a size of $150 \,\mu$m$\times 150\,\mu$m$\times 14\,$nm. As the included Co-atoms lead to a granular structure, with decoupled magnetic grains with a size of $20.6\pm 4\,$nm, we can model the system with discrete cells with a dimension of $(41.2\times 41.2\times 14)\,$nm$^3$. To cover the desired area we need $3640^2$ cells.
Subsequent to the time propagation over $300\sigma_t$, the magnetic configuration is relaxed, which means that the precessional term in the LLG is disregarded in order to achieve a fast approach towards the final stable magnetic configuration. 

The pulse induced ring pattern of the magnetic domains pointing either up or down (with respect to  the easy direction of the magnetic films) is well captured by our micromagnetic simulations.
As pointed out in \cite{27},  the critical precessional angle $\phi\geq\pi/2$ is determined  by the local strength of the magnetic field and indicates  the achieved angular velocity $\omega$. 
The pulse duration $\sigma_t$ plays a crucial role \cite{28a,28b,28c}.  As discussed in Ref.\cite{28a,28b,28c}, an appropriate sequence of ps pulses allows for  an optimal control scheme achieving  a ballistic magnetic switching, even in the presence of high thermal fluctuations.  Longer pulses might drive the system back to the initial state \cite{28a,28b,28c}. So, the critical precessional angle and $\sigma_t$ are the two key parameters  \cite{27} for the established   final precessional angle $\phi=\omega\sigma_t$.  Note,  the demagnetization fields are also relevant, as inferred  from Fig. \ref{fig_comp} but they do not change the main picture.
\subsection{Nanoscale magnetization dynamics}
\label{sec:2.2}
Having substantiated our methods against experiment we turn to the main focus of our study, to be specific the generation of magnetic excitations on the nanoscale. 
One possible application can be found regarding nanowire geometries. Our aim is to excite such magnetic systems in two different ways. Obviously, the creation of domain walls analogously to the CoCrPt-system discussed before should be possible. On top of this, the confined system leads to a unidirectional transport of spin-waves, also called magnons, which are stimulated by the abrupt excitation of the magnetic moments close to the beam. 
This setup becomes interesting for strongly focused beams, that cause magnon-modes beyond the linear-response regime, as the LLG couples the magnetization to the demagnetizing fields, which again act on the magnetization dynamics.

We choose a nanowire with a size of $4000\times 50\times 2\,$nm$^3$, which corresponds to a system of $2000\times 25\times 1$ simulation cells. The material parameters are the same as the ones before adapted from the experiment, except for the Gilbert damping parameter, which is reduced to $\alpha=0.001$ in order to incorporate magnonic excitations. In practice, this can be achieved by preventing the system from exhibiting granular structures. Even if the reduction is not sufficient for this material, the principle can be transferred easily to other thin magnetic films with out-of-plane anisotropy.  The electron beam's duration is $\sigma_t=2.3$ as before, whereas the number of electrons is reduced, but focused on the nanoscale and striking at nanowire's center. 

\begin{figure}
	\begin{center}
		\includegraphics[width=\linewidth]{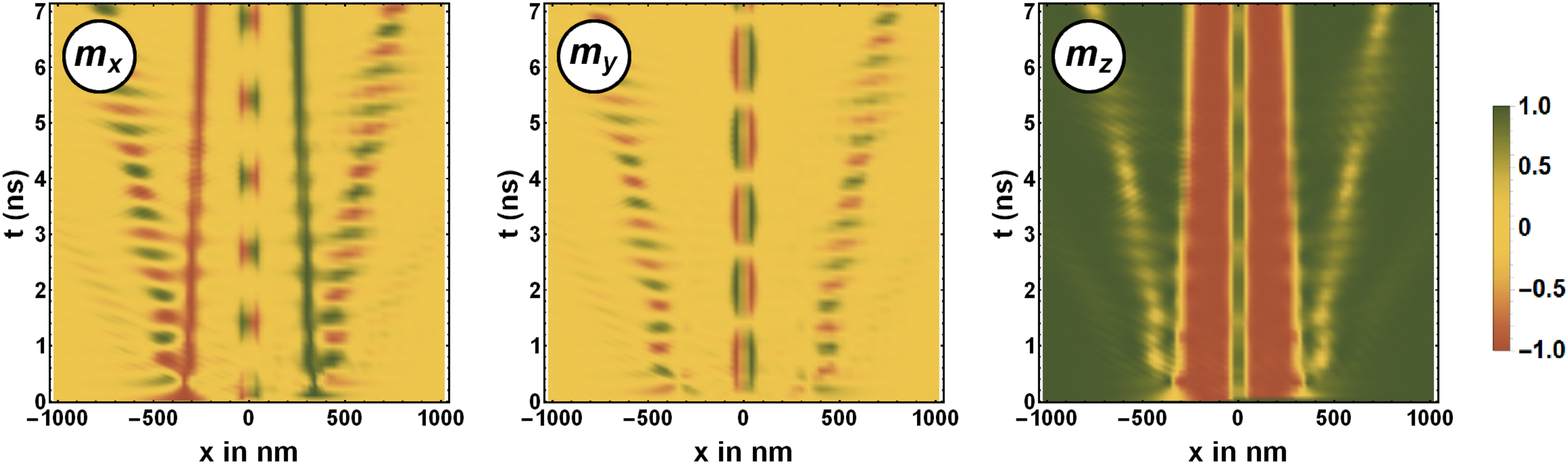}
		\caption{Time propagation of the magnetization's components for a nanowire of  $4000\times 50\times 2\,$nm$^3$ size and electron beam parameters being $\sigma_t=2.3$, $\sigma_{xy}=100\,$nm and $N_e=5\times 10^6$. }
		\label{fig_wire1}
		\includegraphics[width=\linewidth]{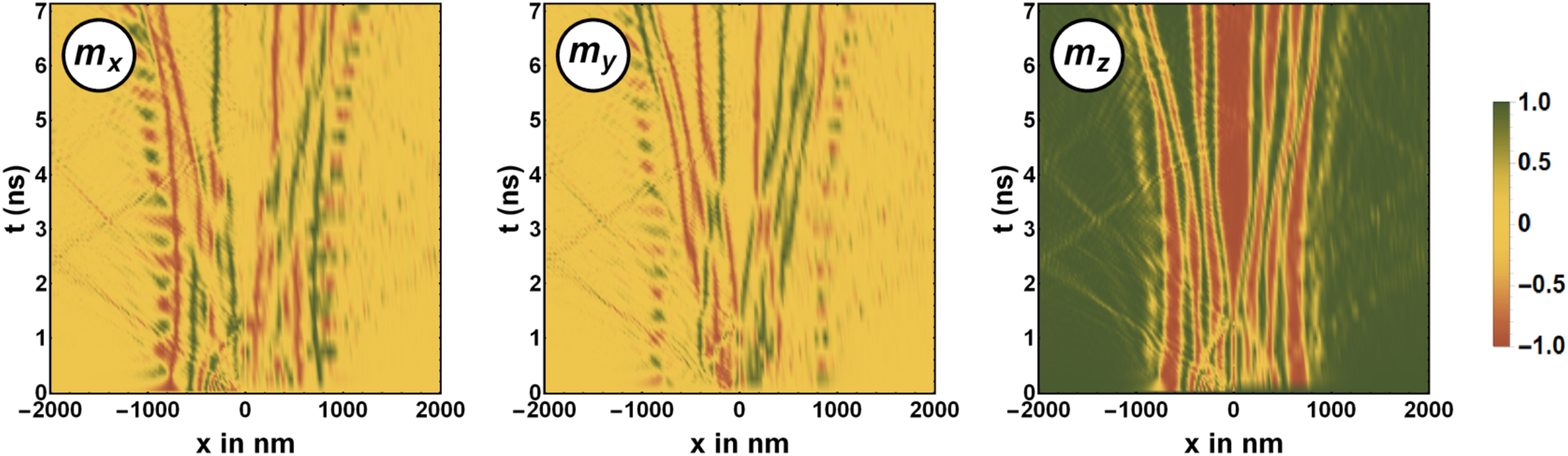}
		\caption{Time propagation of the magnetization's components for a nanowire of  $4000\times 50\times 2\,$nm$^3$ size and electron beam parameters being $\sigma_t=2.3$, $\sigma_{xy}=30\,$nm and $N_e=5\times 10^7$.}
		\label{fig_wire2}
	\end{center}
\end{figure}

Two different examples are shown in \cref{fig_wire1} and \cref{fig_wire2}. The graphics show the magnetization's components' time propagation, averaged over the wire's breadth. 
In \cref{fig_wire1} a number of $N_e=5\times 10^6$ electrons is bundled on a normal distribution with $\sigma_{xy}=100\,$nm. A few things in the magnetization are worth to be mentioned. Most distinctly is the creation of two regions with inverted magnetization near the beam's center ($50\,$nm$\leq |x|\leq 300\,$nm). The outer border regions show a N\'eel type domain wall, whereas the region in between the two domains features a rotating behavior of the in-plane components, which will relax towards a single- or two-domain state on a longer time-scale. 
On top of this two branches of magnons, propagating along the wire are present. One rather fast propagating spin-wave of minor amplitude and one more pronounced magnon possessing a smaller group velocity. 
By increasing the beam's intensity ($\sigma_{xy}=100\,$nm, $N_e=5\times 10^7$) the resulting magnetization dynamics become more complex and multiple branches of magnons interfering with each other, but also domain-walls propagating along the wire can be observed (see \cref{fig_wire2}). Especially rapidly moving magnons, which are reflected by the simulation boxes boundaries are present. Because of the magnetization's non-linear feedback in terms of the demagnetizing fields, a complex pattern occurs. 
This can be used to analyze magnon-excitation in nanostructures beyond linear-response.

\subsection{Imprinting of topological magnetic structures}
\label{sec:2.3}
The generation of topologically  protected magnetic excitations such as skyrmions via pulsed electron pulses is another possible application for the established method. 
A recent work \cite{29} evidences that  ultra thin nanodiscs of materials such Co$_{70.5}$Fe$_{4.5}$Si$_{15}$B$_{10} $\cite{30} sandwiched between Pt  and Ru/Ta  are  well suited for our purpose since they exhibit Dzyaloshinskii-Moriya (DM) spin-orbital coupling.
Hence the magnetization's structure may  nucleate spontaneously into skyrmionic configurations.
We adapted the  experimentally verified parameters for this sample and present results for the magnetic dynamics triggered by short electron beam pulses.

\begin{figure}
	\begin{center}
		\includegraphics[width=0.5\linewidth]{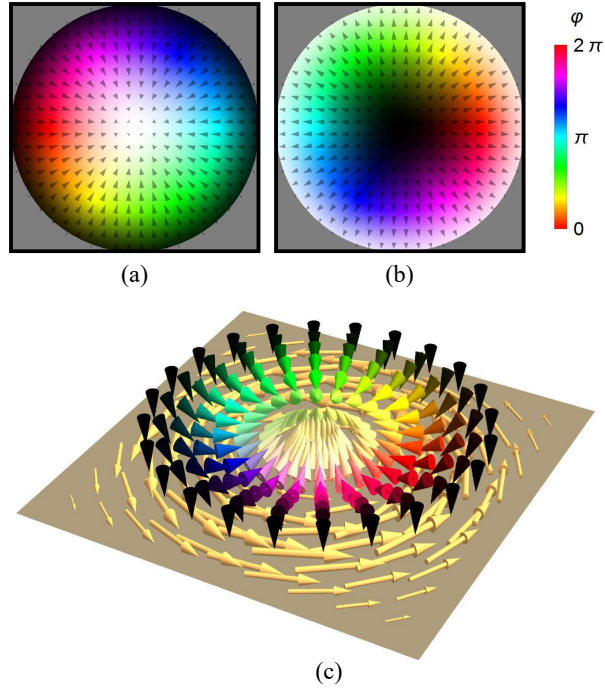}
		\caption{Magnetic ground states for a nanodisc with a diameter of 300\,nm and a thickness of 1.5nm. The material parameters are $M\ind{sat}=450\times 10^3\,$A/m, $A\ind{ex}=10\,$pJ/m, $\alpha=0.01$, $K_u=1.2\times 10^5\,$J/m$^3$ (out-of-plane anisotropy), and the interfacial DMI-constant $D\ind{ind}=0.31\times 10^{-3}\,$mJ/m$^2$. (a) corresponds to $N=1$, whereas (b) possesses $N=-1$, both skyrmions are of the N\'eel type.  Bottom panel illustrates pictorially  the influence of the magnetic field associated with the electron bunch. The cones correspond to the initial magnetic configuration as in (a) and (b), whereas the golden arrows show the induced magnetic field. The resulting torque points perpendicular to the magnetization, affecting the magnetic configuration accordingly. }
		\label{fig_groundstate}
	\end{center}
\end{figure}

Taking a nanodisc of a variable size the ground state with a topological number $|N|=1$ is realized after propagating an initially  homogeneous magnetization in $\pm z$ direction according to the Landau-Lifshitz-Gilbert equation (LLG) including DM interactions \cite{34,35,36,37}.
The two possible ground states, depending on the initial magnetization's direction are shown in \cref{fig_groundstate} along with the material's parameters.

Our main focus is on how to efficiently and swiftly switch between these skyrmion states via a nano-focused relativistic electron pulse, an issue of relevance when it comes to fundamental research or practical applications like data storage.
While currently such pulses can be generated with micron size beam dimensions \cite{ued_ref} future sources are expected to reach focus sizes down to the few nm range \cite{32}. In principle the possibility of beam damage occurring in the beam’s focus as in the case of the experiment in ref.\cite{27} is present. However, ongoing experiments with relativistic electron beams \cite{ued_ref} indicate that the use of ultrathin freestanding films may alleviate damage concerns. 

\begin{figure}[th]
	\begin{center}
		\includegraphics[width=0.6\linewidth]{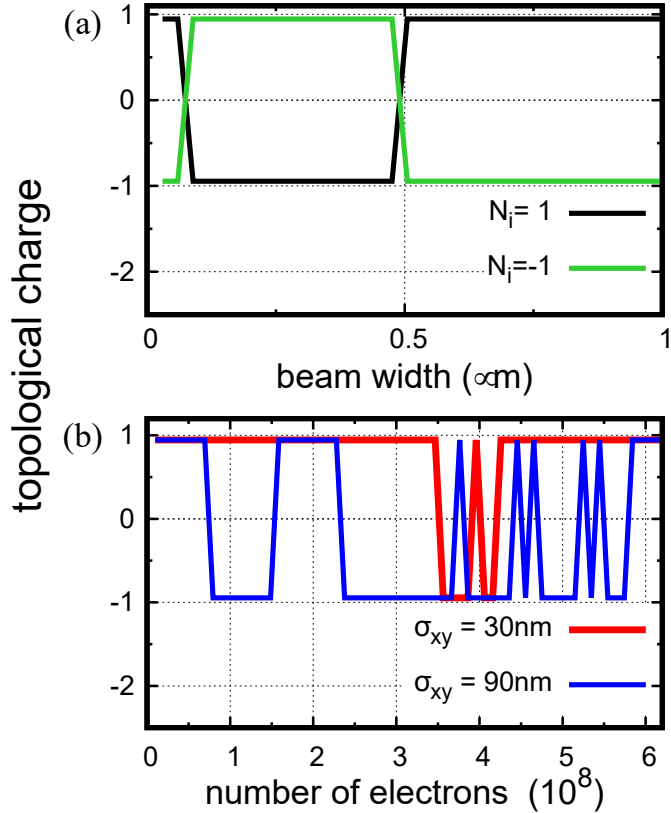}
		\caption{Varying the number of electrons per pulse or the spatial enlargement of the pulse, the imprinted  topological charge  can be tuned. The pulse duration is set to $2.3\,$ps. Black and green curves correspond respectively to starting with a magnetic ordering having $+1$ or $-1$ topological charge, as shown in \cref{fig_groundstate} for different pulse widths. Both the blue and red curve start from $N_i=+1$.
			The sample is a magnetic disc (diameter $d=300\,$nm) which is  irradiated  with a Gaussian beam pulse with $\sigma_{xy}=30\,$nm (and $90\,$nm) in case of the bottom graph, respectively the upper graphs' beam has a constant number of $n_e=10^8$ electrons. }
		\label{fig_dur}
	\end{center}
\end{figure}

Topologically protected magnetic configurations, like magnetic skyrmions, are well defined quasiparticles. They can be characterized mathematically by the topological number $N=\frac{1}{4\pi}\int \vec{m}\cdot\left(\frac{\partial \vec{m}}{\partial x}\times\frac{\partial \vec{m}}{\partial y}\right)\dd x\dd y$\cite{33} also called winding number, which counts how often the unit vector of the magnetization wraps the unit sphere when integrated over the two-dimensional sample.
Therefore, skyrmions are typically a quasiparticle in thin (mono)layers. The topological number  adopts integer values indicating the magnetic configuration to be skyrmionic ($N=\pm 1$) or skyrmion multiplexes ($|N| >1$). If the topological number is not an integer the topological protection is lifted and the magnetic texture is unstable upon  small perturbations.
The topological stability of skyrmionic states stem from the necessity of flipping at least one single magnetic moment by $180^\circ$, to overcome the barrier and transfer the state into a "trivial" state, like a single domain or vortex texture.
In the following, we will attempt to overcome this topological energy barrier with a magnetic "kick" so that the magnetization will be converted into a state of different topological invariant.
Advantageous is the spatial structure of the magnetic field curling around the beam's center, which gives a good point of action in order to manipulate topologically protected configurations.\\
For magnetic systems, a minimum time of exposure is necessary, whereas the spatial focus of the beam is limited. To overcome this conflict, the pulse duration is fixed at $2.3\,$ps as before, when nothing different is mentioned. Starting from such an electron beam two main parameters can be adjusted to achieve the favored reaction of the nanodiscs. Those are the pulse width and the number of electrons, which will be treated independently. 	In \cref{fig_dur}, the final topological charges after a single Gaussian electron pulse irradiating a nanodisc are plotted as a function of the  number of electrons and the width of the Gaussian distributed electrons. 
The results do not show the transient  time evolution of the sample but only the final steady-state  values of the winding number. They are obtained by applying an electron pulse, propagating the magnetization during the pulse, and relaxing the magnetic configuration afterward as to  approach a local minimum of the free energy's hypersurface.
We note  the strong correlation between the change of the topological charge and the number of electrons or accordingly the beam width. Relatively large intervals of both parameters lead to the same final values for $N$. We note that not only the variation of these control parameters, but also of the duration of the pulse is experimentally accessible, particularly in a nano-apex ultrafast transmission electron microscope \cite{5,6,7,8,9,10,11,12,13,14,15,16,17,18,19,20,21,22}.
Noteworthy,  the graphs for opposite initial configurations (see \cref{fig_dur}(a)) are axially symmetric with respect to the $x$ axis. This can be explained by the coinciding symmetry centers of the pulse and the skyrmionic structure.
This symmetric and robust behavior can be exploited to switch between the accessible different values for the topological charges which are quite close to the ideal integer values that would be realized in an infinitely small discretization.

Interestingly, the switching between the two stable states occurs repetitively for an increase in the number of electrons, whereas the spatial manipulation of the beam leads to one regime only in which the fields are sufficient to switch the topological number. The first observation can be explained with the schematics shown in fig.\ref{fig_groundstate}c). Depending on the strength of the pulse the magnetic moments are forced to rotate multiple times around the $\hat{e}_\varphi$ vector in a collective manner, as each moment of equal distance to the center experiences the same torque. The final position of the surrounding moments couples backward to the center and determines the new topological charge. The electron number linearly translates to the peak magnetic field, whereas the beam width has a more complicated influence. 
When the width is increased the spatial profile in the $xy$-plane is manipulated, as the maximum magnetic field is shifted towards the disc's rim and beyond. How the system reacts on this changes depends crucially on the exact profile of the beam, especially on the point of maximum magnetic field strength, as can be seen in \cref{fig_dur}(a).
This leads to the question of the optimum parameter regime, to manipulate the system reliably, which can not finally be answered as it strongly depends on the experimentally available capabilities. Hence this work focuses on an exemplary study on the effect.

The same switching phenomenon as discussed before can also be observed for different setups. Weaker pulses, as long as they are able to overcome the internal fields to excite the system, can be used as well, but obviously, the field's amplitude translates to the strength of the resulting torque. This implies a longer radiation time needed for pulses of lower intensity to be capable of switching the system.	

In the case of different materials or geometries, the accessible topological states have to be investigated, before they can be utilized. Otherwise undesired, interstitial states might be achieved by accident and the switching is not deterministic anymore. 
\section{Summary and Outlook}
We have shown using micromagnetic calculations that relativistic electron pulses of few ps duration can indeed induce magnetization dynamics in ferromagnetic nanostructures and extended films as observed experimentally \cite{27}. Contrary to micron sized electron beams employed experimentally so far \cite{27}, nanofocusing allows us to reduce the number of electrons dramatically in order to achieve magnetic switching. We demonstrated this for the case of ferromagnetic nanowires where the nanofocused electron pulses enabled magnetic switching on one end of the wire with magnon excitations propagating along the wire towards the other end. We also predict a novel way of switching the winding sense of magnetic skyrmions using the tangential fields of electron pulses focused down to the skyrmion size. We predict that such magnon excitations can be driven with electron numbers as low as $10^6$ electrons/pulse. This is within reach at present electron diffraction experiments \cite{ued_ref}. Attempts at achieving the required nanofocus at such sources are currently underway. We note that the predicted magnetization dynamics could be imaged using VUV \cite{44} and soft x-ray photons \cite{45}. Since both ultrafast electron diffraction and imaging experiments start with laser-generated photoelectrons the synchronization of fs probe laser systems with the electron pulses is straightforward \cite{ued_ref}. The necessary probing of magnetization dynamics has recently been demonstrated using circularly polarized photons for high-harmonic laser sources \cite{46}. 


\section*{Acknowledgments}
A. F. S. and J. B. are supported by the German Research Foundation (No. SFB 762) and the Priority Programme 1840. H.A.D. acknowledges support by the U.S. Department of Energy,  Office of Basic Energy Sciences, Materials Sciences and Engineering Division under Contract No. DE-AC02-76SF00515.








\end{document}